\documentclass[acus]{JAC2003}

\usepackage{graphicx}
\usepackage{booktabs}
\usepackage{amssymb}
\usepackage{cite}
\begin{document}
\title{TOWARDS ZEPTOSECOND-SCALE PULSES FROM X-RAY FREE-ELECTRON LASERS}

\author{D.J. Dunning, N.R. Thompson, ASTeC, STFC Daresbury Laboratory and Cockcroft Institute, UK\\
B.W.J. McNeil, Department of Physics, SUPA, University of Strathclyde, Glasgow, UK}

\maketitle

\begin{abstract}
The short wavelength and high peak power of the present generation of free-electron lasers (FELs) opens the possibility of ultra-short pulses even surpassing the present (tens to hundreds of attoseconds) capabilities of other light sources – but only if x-ray FELs can be made to generate pulses consisting of just a few optical cycles. For hard x-ray operation ($\lesssim$0.1nm), this corresponds to durations of approximately a single attosecond, and below into the zeptosecond scale. This talk will describe a novel method~\cite{mlab} to generate trains of few-cycle pulses, at GW peak powers, from existing x-ray FEL facilities by using a relatively short `afterburner'. Such pulses would enhance research opportunity in atomic dynamics and push capability towards the investigation of electronic-nuclear and nuclear dynamics. The corresponding multi-colour spectral output, with a bandwidth envelope increased by up to two orders of magnitudes over SASE, also has potential applications.
\end{abstract}

\section{Introduction}
\label{main}

The motivation for generating short pulses of light is to study and influence ultra-fast dynamic processes. To do this, radiation pulses on a shorter scale than the dynamics involved are required. The timescales of different processes have been described by Krausz and Ivanov~\cite{atto2}: Atomic motion on molecular scales occurs at femtosecond ($10^{-15}$~s) to picosecond ($10^{-12}$~s) scales, electron motion in outer shells of atoms takes place on tens to hundreds of attoseconds, and electron motion in inner shells of atoms is expected to occur around the scale of a single attosecond ($10^{-18}$~s). At faster scales still are nuclear dynamics, which are predicted to occur at zeptosecond ($10^{-21}$~s) time scales .

The record for the shortest pulse of light has seen a progression from approximately 10~ps in the 1960s to around 67~attoseconds generated recently by Chang et al.~\cite{shortestpulse}---a development of approximately five orders of magnitude in five decades. As noted by Corkum et al.~\cite{atto3, atto1}, it is particularly relevant to consider the way in which this frontier progressed. The duration of a pulse of light is its wavelength, $\lambda_r$ multiplied by the number of optical cycles, $N$, divided by the speed of light. Initially progress was made in conventional lasers operating at approximately a fixed wavelength ($\lambda_r\approx$600~nm), by reducing the number of optical cycles. This continued until, in the mid 1980s, pulses of only a few cycles could be generated (corresponding to a few fs), then could proceed little more.

It took a transformative step---high harmonic generation (HHG)~\cite{atto2, atto3, atto1}, for progress to continue by (in very simple terms) reducing the wavelength of the generated light. This technology allowed pulses in the attosecond scale to be generated for the first time, and now reaches just under a hundred attoseconds.

It seems that a further step to shorter wavelength is now required to progress to significantly shorter pulses. Proposals are being developed outlining how future progress in HHG might achieve this~\cite{zepto_HHG}. Alternatively, x-ray free-electron lasers (FELs) (reviewed in several recent papers~\cite{barletta, natphoton, braun}) presently surpass HHG sources in terms of shortest wavelength by approximately two orders of magnitude, and it is this property which first suggests FELs as a promising candidate for progressing to shorter radiation pulses than are available today.

%Here introduce the paper, and put a nome¬nclature if necessary, in a box with the same font size as the rest of the paper. The paragraphs continue from here and are only separated by headings, subheadings, images and formulae. The section headings are arranged by numbers, bold and 10 pt. Here follows further instructions for authors.

%\begin{nomenclature}
%\begin{deflist}[A]
%\defitem{A}\defterm{radius of}
%\defitem{B}\defterm{position of}
%\defitem{C}\defterm{further nomenclature continues down the page inside the text box\vspace*{-8pt}}
%\end{deflist}
%\end{nomenclature}
%\vspace*{8pt}

\section{Short-pulse potential of free-electron lasers}
The free-electron laser in fact has two particular advantages which give it potential for pushing the frontier of short pulse generation. The first, as described in the previous section, is short wavelength. Recent new FEL facilities (LCLS~\cite{LCLS} commissioned in 2009, and SACLA~\cite{SACLA} commissioned in 2011) have extended FEL operation down to approximately 0.1~nm. Assuming that pulses of only a few optical cycles could be attained, this would correspond to pulse durations of approximately a single attosecond---approximately two orders of magnitude shorter than present HHG sources, and four orders of magnitude beyond conventional lasers.

Of course x-ray sources other than FELs have been available for many years, however the peak powers are insufficient to deliver a significant number of photons within an attosecond timescale.
It is the high peak power of the free-electron laser (exceeding synchrotrons - the next highest intensity source of x-rays - by approximately 9 orders of magnitude) which gives it potential to push the frontier of ultra-short pulse generation. A hard x-ray FEL typically generating approximately 20~GW peak power, corresonds to $10^{25}$ photons/second. For a pulse duration of a single attosecond this would correspond to $10^{7}$ photons per pulse.

The challenge for reaching the very shortest pulses from FELs - as described in the following sections - will be to minimise the number of cycles per pulse.

\begin{figure*}[t!]
    \centering
    \includegraphics*[width=100mm]{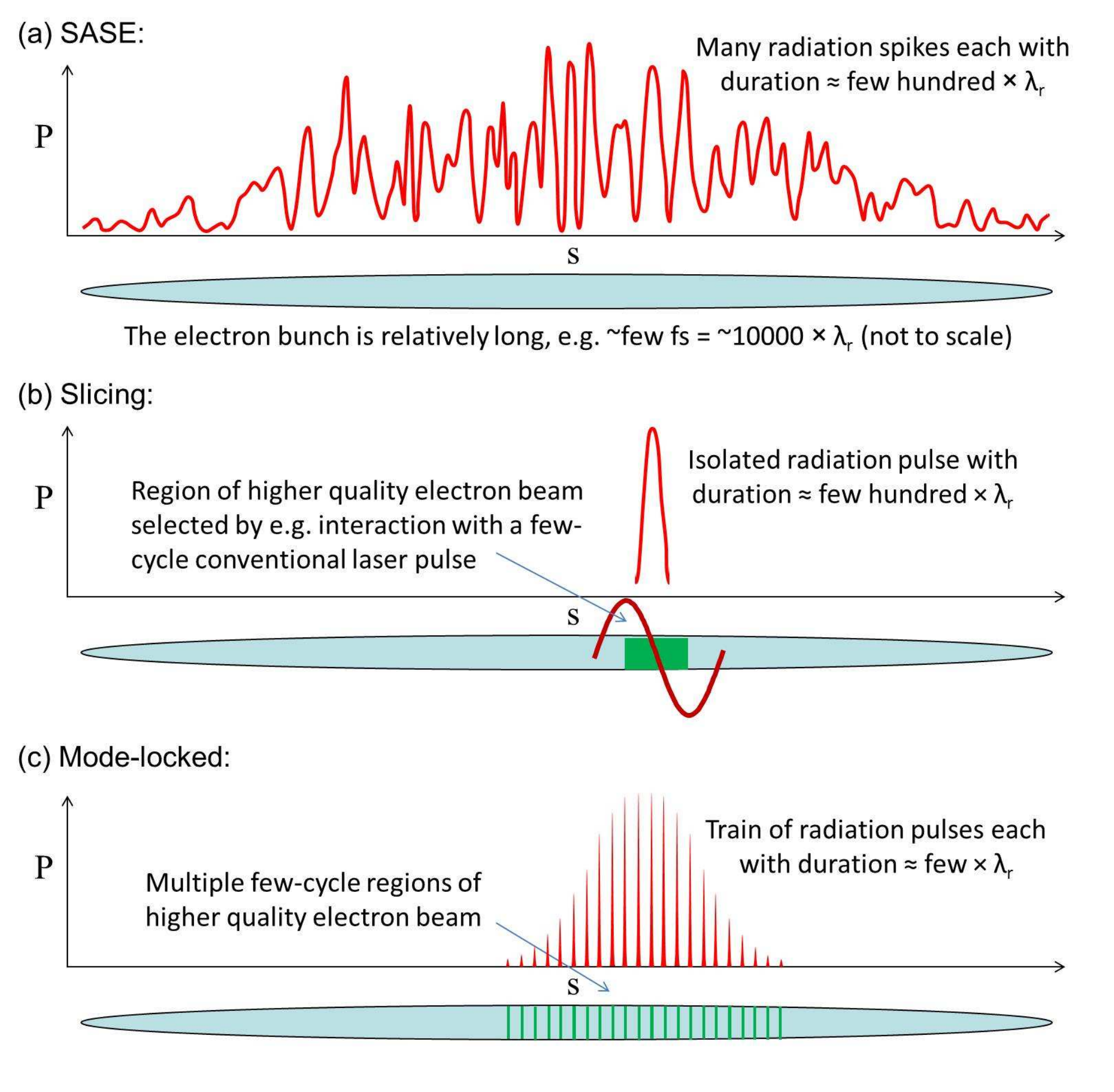}
    \caption{Figure to illustrate different concepts for FEL operation (not to scale): (a) Typical hard x-ray SASE FEL output consists of a number of radiation spikes, each of length $\approx l_c$ (a few hundred optical cycles); (b) Example of proposals to `slice' the electron bunch such that a single pulse of length $\approx l_c$ (a few hundred optical cycles) is generated; (c) The mode-locked FEL concepts work by slicing the electron bunch into regions $\ll l_c$, and periodically shifting the radiation to generate a pulse train with pulses on a similarly short scale.}
    \label{SASE_slicing_ML}
\end{figure*}

\section{Standard operating mode of a hard x-ray FEL - SASE}
Present hard x-ray FELs normally operate in the high-gain amplifier mode generating self-amplified spontaneous emission (SASE) (as described by Bonifacio et al.~\cite{bnp}), which has noisy temporal and spectral properties~\cite{bonprl}. A relativistic electron bunch is injected into a long undulator (an alternating polarity magnetic field with period $\lambda_u$) which causes the electrons to oscillate transversely and so emit radiation. The electrons' transverse oscillation allows a resonant, co-operative interaction with the co-propagating radiation field of resonant wavelength $\lambda_r=\lambda_u(1+\bar{a}_u^2)/2\gamma_0^2$~\cite{natphoton}, where $\bar{a}_u$ is the rms undulator parameter and $\gamma_0$ is the mean electron energy in units of the electron rest mass energy.

The co-operative instability results in an exponential amplification of both the resonant radiation intensity and the electron micro-bunching, $b=\langle e^{-i\theta_j}\rangle$~\cite{bnp}, where $\theta_j$ is the ponderomotive phase~\cite{natphoton} of the $j^\mathrm{th}$ electron.
In the one-dimensional limit, the length-scale of the exponential gain is determined by the gain length $l_g=\lambda_u/4\pi\rho$, where $\rho$ is the FEL coupling parameter~\cite{bnp} (typically $\rho\approx10^{-4}-10^{-3}$ for x-ray FELs). The exponential growth saturates when a fraction approximately equal to $\rho$ of the electron beam power is extracted into radiation power.
In the undulator, a resonant radiation wavefront propagates ahead through the electron bunch at a rate of one radiation wavelength, $\lambda_r$ per undulator period, $\lambda_u$. This relative propagation, or `slippage' in one gain length $l_{g}$ is called the `co-operation length', $l_{c}=\lambda_r/4\pi\rho$~\cite{bmp}, which determines the phase coherence length.

The electron bunch is relatively long, at least in the context of this paper, with a few-fs bunch corresponding to $\approx10^{4}\times\lambda_r$ at 0.1~nm. The total duration of the radiation emission is similar to that of the electron bunch length (e.g. a few fs), however it consists of sharp spikes on the much shorter scale of the co-operation length, typically a few hundred radiation wavelengths, corresponding to approximately 100~as for hard x-ray FELs, as shown in Figure~\ref{SASE_slicing_ML}~(a).

\subsection{Slicing a Single SASE Spike}

Since each SASE spike acts independently it has been proposed by a number of groups e.g.~\cite{bonprl,saldin,zholentsNJP,emma,ml_esase} that only one spike can be made to occur, either by reducing the bunch length or by 'slicing' the electron beam quality. Experimental progress has been made for a few of these methods, including reducing the electron bunch length~\cite{LCLS_single_spike_SASE}, and by slicing part of the beam via emittance spoiling~\cite{emma, LCLS_spoiling}. A short-pulse technique using chirped electron beams and a tapered undulator has been demonstrated at visible wavelengths~\cite{Giannessi_PRL_11, Marcus_APL_12} and could be extended to x-ray.

\begin{figure*}[hbt]\vspace*{0pt}
%\centerline{\includegraphics{fx1}\hspace*{5mm}\includegraphics{fx1}}
\centering
\includegraphics*[width=120mm, clip=]{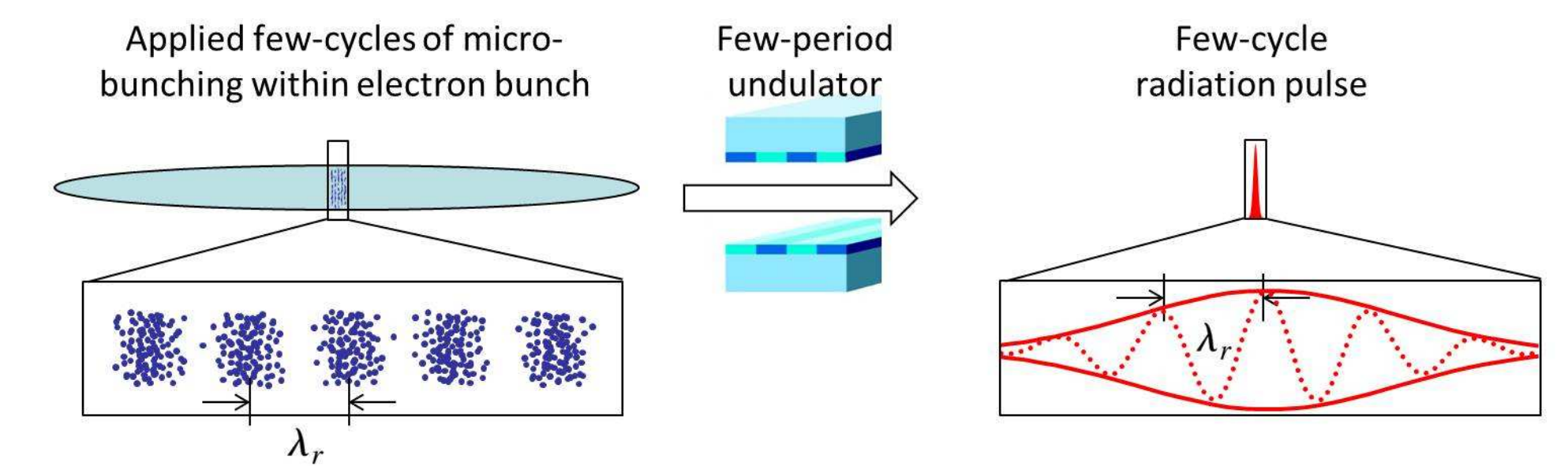}
\caption{Figure to illustrate a concept to generate an isolated few-cycle pulse from an electron bunch. Instead of using the FEL interaction, an external source is used to induce microbunching over a region only a few cycles in length, followed by a few-period undulator to emit a few-cycle pulse.}
\label{Induced_microbunching}
\end{figure*}
%\vspace{-10pt}

A number of proposals (e.g. by Saldin et al.~\cite{saldin}) suggest using a few-cycle conventional laser pulse to pick out part of the electron beam, as shown in Figure~\ref{SASE_slicing_ML}~(b), which has the advantage of generating a radiation pulse synchronised to the external source.
Picking out one SASE spike for present hard x-ray FEL parameters corresponds to a few hundred optical cycles or $\approx$100~as, which would be close to the frontier presently set by HHG, and at shorter wavelength and higher photon flux (methods are predicted to reach normal SASE saturation power levels or even higher). This technique therefore has exceptional potential, however in terms of shortest pulse duration there is still potential for a further two orders of magnitude reduction by reducing the number of cycles per pulse. Possible methods of doing this are considered in the following sections.

\section{Issues in generating few-cycle pulses from FEL amplifiers}
Exponential amplification of the radiation power in an FEL amplifier requires a sustained interaction between the radiation field and the electron bunch. This presents a difficulty for generating few-cycle radiation pulses from FELs, since the slippage of the radiation relative to the electrons means that a few-cycle radiation pulse can only interact with a fixed point in the electron beam for a few undulator periods before slipping ahead of it.
For example if we were to use one of the methods described in the previous section but were to slice a high quality section of the electron beam much shorter than one SASE spike (e.g. a few cycles) then the rapid slippage of the generated radiation ahead of the high quality region would significantly inhibit FEL amplification~\cite{bonprl,limit}. Future increases in electron beam brightness may reduce the FEL co-operation length and so enable reducing pulse durations from this method to some extent.

Another route forward may be the superradiant regime in a seeded FEL amplifier, which has been addressed in theory~\cite{SR1,SR2}, and in experiment, with short pulses generation observed in both direct seeding~\cite{SR3, SR4} and harmonic cascade~\cite{SR5, SR6} configurations. In such techniques a short section of an electron bunch is seeded such that it reaches saturation before the rest of the bunch (which starts up from noise). Beyond saturation the FEL interaction proceeds into the superradiant regime in which the radiation intensity continues to increase (though quadratically with distance through the undulator, $z$, rather than exponentially), and the pulse length reduces as $z^{-1/2}$. Compared to the exponential regime, where the `centre-of-mass' of a radiation spike is kept close to the electron longitudinal velocity due to amplification, in this mode the radiation pulse propagates closer to the speed of light, so forward relative to the electrons. Consequently it propagates into `fresh' electrons, provided the rest of the bunch starting from noise has not reached saturation. This technique has been demonstrated at longer wavelengths~\cite{SR3, SR4, SR5, SR6}, however its scalability to hard x-ray wavelengths still requires significant development. Few-cycle pulses have been attained via superradiance in FEL oscillators, but FEL oscillators operating at x-ray wavelengths are still under development~\cite{xfelo}, and present ideas for suitable mirror cavities have very narrow bandwidth which would seem incompatible with the broad-band operation required for few-cycle pulses.

An alternative might be to disregard FEL amplification to establish microbunching, and instead use an external source (though this itself may be a FEL) to induce microbunching (or a single sharp current spike) over a region only a few cycles in length and then make it radiate in an undulator, as shown in Figure~\ref{Induced_microbunching}. There are several proposals to do this, such as by Zholents and Fawley~\cite{zholents} or by Xiang et al.~\cite{echo_atto}, though again the slippage has a limiting effect. If the number of undulator periods in the radiator is greater than the number of cycles in the microbunched region, then the slippage effect dominates and lengthens the pulse. The undulator must therefore be similarly short - also a few periods - otherwise slippage of the radiation relative to the electrons broadens the pulse. As a consequence proposals for this type of technique predict relatively low power compared to FEL saturation, however the power could potentially be increased by future improvements in electron beam brightness. Requiring the microbunching to be imposed by an external source may also present difficulties in scaling such techniques to the shortest wavelengths of FELs in some cases.

\section{Few-cycle pulses via pulse-train operation}
\begin{figure*}[hbt]\vspace*{4pt}
%\centerline{\includegraphics{fx1}\hspace*{5mm}\includegraphics{fx1}}
\centering
\includegraphics*[width=120mm]{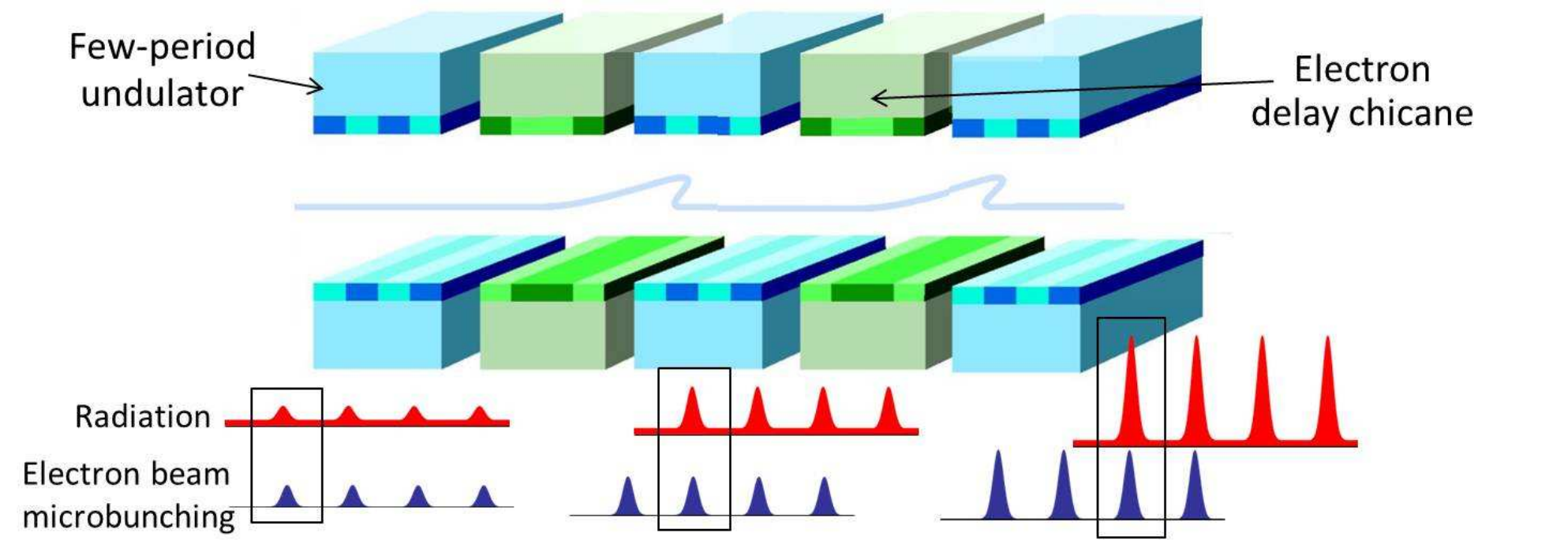}
\caption{Schematic layout of a section of the undulator used in the mode-locked FEL amplifier and mode-locked afterburner techniques. Chicanes are used to periodically delay the electrons to keep the developing radiation spikes overlapped with regions of high microbunching.}
\label{ML_schematic}
\end{figure*}

We noted in the previous section that the slippage of the radiation ahead of the electrons seems to imply a trade-off between maximising the emitted power (requiring a long interaction), and minimising the pulse duration (requiring a short interaction). However the mode-locked FEL amplifier concept proposed by Thompson and McNeil~\cite{mlsase} circumvents this by dividing the long FEL interaction into a series of short interactions separated by a longitudinal re-alignment of the radiation and electron beam. Such re-alignment can be achieved via magnetic chicanes to delay the electron beam relative to the radiation, as shown in Figure~\ref{ML_schematic}.

In this case the radiation always propagates ahead relative to the electrons, which is not advantageous for amplifying an isolated ultra-short pulse. However, it allows a \emph{train} of ultra-short pulses to be amplified. If we consider one pulse in the train, it interacts with the electron beam for several undulator periods, increasing the intensity of the radiation and increasing the microbunching of the electron beam over that region. It then is shifted forward relative to the electrons, to interact with the region of the electron beam previously aligned with the radiation pulse preceding it in the train. This allows a series of short interactions, thereby allowing the pulse duration to be minimised while at the same time maximising the power.

Though such a technique would allow an external pulse train source (such as that available from HHG) to be amplified~\cite{mlseeded}, such sources are not available at hard x-ray wavelengths so the FEL starts up from noise. In this case multiple interleaved pulse trains may be supported, and a comb structure variation must be applied to the electron beam properties~\cite{mlsase, s2e_mlfel, njp_ml_curmod} to select a single clean pulse train structure, as shown in Figure~\ref{SASE_slicing_ML}~(c).
The minimum number of optical cycles per pulse from this method is approximately the number of undulator periods per section, so it could potentially deliver few-cycle pulses. However, this would require significantly modifying existing FELs, which are typically divided into modules of several hundred periods.

\section{New concept for few-cycle pulses}
A new method has recently been developed by the authors~\cite{mlab} that would allow existing x-ray FEL facilities to generate trains of few-cycle radiation pulses via the addition of only a relatively short `afterburner' extension that could relatively easily be added to existing facilities.

\subsection{Slicing Multiple Few-cycle Regions}
The technique involves preparing an electron beam with periodic regions of high beam quality, each region of length $\ll l_{c}$, prior to injection into a normal FEL amplifier. In isolation, each region would be insufficient to support FEL amplification, however by positioning a number of these regions reasonably closely spaced, they can interact via the radiation propagating between them - and support amplification.
This variation in electron beam quality can be achieved for example via electron beam energy modulation, where the extrema have less energy spread and are able to lase more easily. Alternatively a current or emittance modulation could potentially be used.
Slippage occurring in the amplifier washes out any short-scale structure in the radiation, however only the high quality regions of the electron beam undergo a strong FEL interaction such that a periodic comb structure is generated in the FEL-induced micro-bunching.

Two methods have been developed to use the comb structure in the electron beam to generate a train of few-cycle pulses. The first is very simple - consisting of just a single few-period undulator - but generates relatively low power. The second method would allow the peak power to approach normal FEL saturation levels.

\subsection{Simple Single Undulator Method}
This method involves blocking the radiation from the amplifier stage, and passing only the micro-bunched electron beam through a single short undulator so that it emits a train of few-cycle radiation pulses~\cite{FEL10_paper}. In this aspect the scheme is similar to a class of methods already described~\cite{zholents, echo_atto}, and is similarly predicted to generate relatively low power. However its simplicity makes it a very promising option for a proof-of-principle experiment, as it requires minimal modification of existing facilities.

\begin{figure}[htb]
   \centering
   \includegraphics[width=\columnwidth]{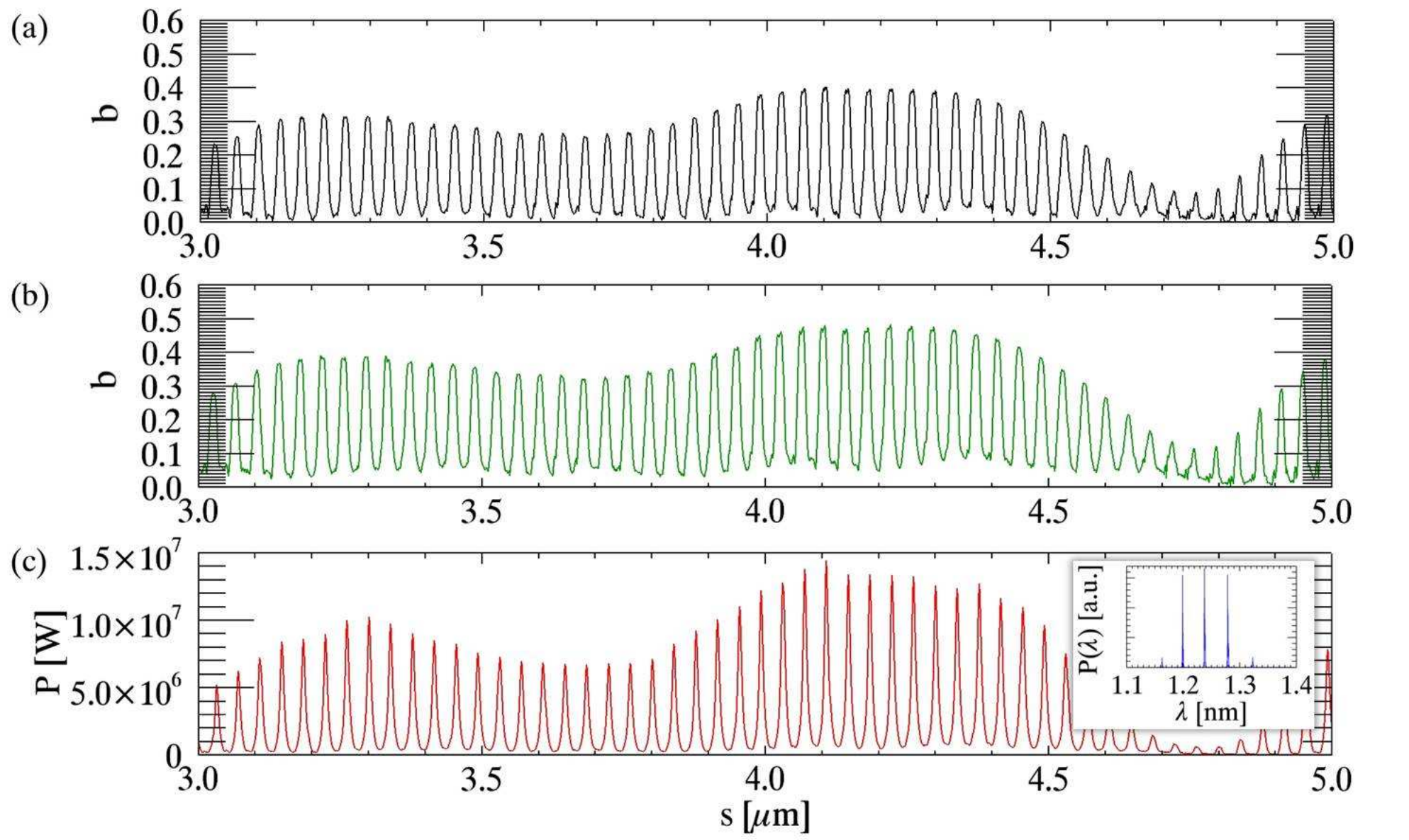}
   \caption{Soft x-ray single short undulator simulation results: Micro-bunching profile at (a) the exit of the amplifier stage and (b) after separation of radiation and electron beam in the chicane, at the entrance to the short undulator; (c) Radiation power profile and spectrum at the end of the 10-period undulator. The duration of an individual pulse is $\sim$10~as~rms.}
   \label{SingleMod_ML_AB_output}
%\vspace{-0.2cm}
\end{figure}

A chicane is used to allow the electron beam and radiation to be separated. Figure~\ref{SingleMod_ML_AB_output} shows the electron beam micro-bunching structure at the exit of the amplification stage, and at the entrance of the single undulator for an example case of a soft x-ray FEL. Also shown in Figure~\ref{SingleMod_ML_AB_output} is the radiation profile and spectrum at the exit of the radiator. Pulses of only five optical cycles are generated - corresponding to 10~as rms in this example. The radiation spectrum consists of discrete modes, with bandwidth increased by up to two orders of magnitude over SASE. Further results are given in~\cite{FEL10_paper}.
%\vspace{-0.2cm}

\subsection{Mode-locked Afterburner}
The second method is more complex but would allow the peak power to approach normal FEL saturation levels.
Once the micro-bunching comb is sufficiently well developed in the amplifier section, but before FEL saturation, the electron beam is injected into a `mode-locked afterburner', which maps the comb structure of the electron micro-bunching into a similar comb of the radiation intensity.
The afterburner comprises a series of few-period undulator modules separated by electron delay chicanes similar to that used in the mode-locked amplifier FEL~\cite{mlsase}, as shown in Figure~\ref{ML_schematic}.
These undulator-chicane modules maintain an overlap between the comb of bunching electrons and the developing radiation comb, each pulse of length $\ll l_c$, allowing it to grow exponentially in power towards FEL saturation power levels ($\gtrsim$GW). The pulses are delivered in trains, since amplification occurs over a number of afterburner modules, and would be naturally synchronised to the modulating laser.

\begin{figure*}[t]
   \centering
   \includegraphics*[width=0.9\textwidth]{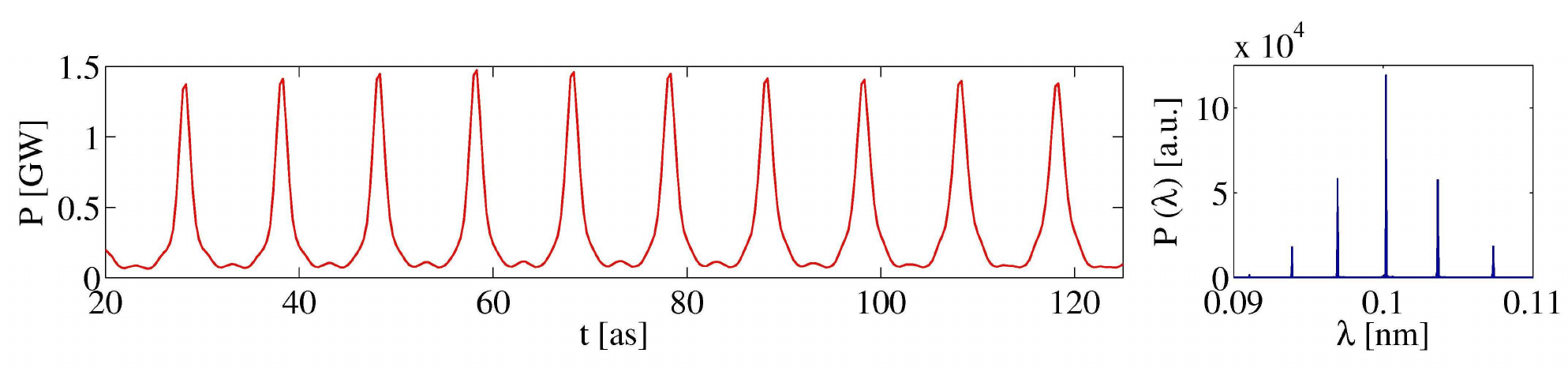}
  % \vspace{-2mm}
   \caption{Hard x-ray mode-locked afterburner simulation results: Radiation power profile (left) and spectrum (right) after 40 modules. The duration of an individual pulse is $\sim$700~zs~rms.}
   \label{0p1nm_ML_AB}
%\vspace{-6mm}
\end{figure*}

Because this method requires no modification of the main undulator we are free to choose the parameters of the afterburner for minimum pulse duration – i.e. short (few-period) undulators. Modelling of the concept at hard x-ray wavelengths, and with 8-period undulators in the afterburner predicts pulse durations of only five optical cycles FWHM, corresponding to 700~zeptoseconds RMS pulse duration~\cite{mlab}. It is also relevant to note that the corresponding spectral properties, which are multi-chromatic within a very broad ($\approx10\%$) bandwidth envelope, may also be useful for some applications.
Figure~\ref{0p1nm_ML_AB} plots the radiation power and spectrum after 40 undulator-chicane modules. In this case the total afterburner consists of 40 modules each consisting of an undulator of length 0.144~m and a chicane of length 0.2~m to give 13.8~m in total.

If the above results are scaled to higher photon energies, e.g to the 50~keV of the proposed x-ray FEL of~\cite{marie}, then pulse durations of 140~zs rms may become feasible~\cite{mlab}.

\section{Conclusion}
There is much potential for generating ultra-short pulses from FELs, and a number of ideas have been proposed. The concepts can be loosely grouped into several categories, each of which has respective merits. Slicing a single SASE spike is predicted to generate isolated pulses with durations of several hundred radiation wavelengths, and high peak power. Using an external source to impart microbunching over a few-cycle region of the beam is predicted to give shorter pulses but relatively low power.

A recent proposal by the authors is predicted to generate few-cycle pulses with high powers by generating the pulses in a train. In frequency space, this pulse train forms a set of phase locked, equally spaced frequency modes, analogous to the modes developed in a conventional cavity mode-locked laser~\cite{mlab, mlsase}. When applied at hard x-ray wavelengths this method predicts individual pulses below the cooperation length scale into the few attosecond, and beyond, into the zeptosecond regime.

A method that may further shorten the individual pulses is to expand the number of modes that are amplified in frequency space - the greater the number of modes, the shorter the pulses. This may be achieved by alternately changing the resonant frequency of each undulator module in the schematic of Fig.~\ref{ML_schematic} to be resonant at a different mode frequency - essentially a `multi-colour' FEL. While the sequencing of the undulator modules resonant at different modes and the extent to which the number of amplified modes can be expanded has yet to be investigated, the authors are unaware of any obvious reasons why the method cannot be extended in this way.

It is anticipated that further significant development in all of the above methods can be achieved and we look forward to future progress.

\end{document}